\documentclass[12pt]{article}
\usepackage{graphicx}
\def \bea{\begin{eqnarray}}
\def \beq{\begin{equation}}

\def \eea{\end{eqnarray}}
\def \eeq{\end{equation}}

\def \od{\overline{D}^0}

\begin{document}

\begin{flushright}
EFI 03-07 \\
hep-ph/0303117 \\
March 2003 \\
\end{flushright}

\bigskip
\medskip
\begin{center}
\large
{\bf Measuring the Relative Strong Phase} \\
{\bf in $D^0 \to K^{*+} K^-$ and $D^0 \to K^{*-} K^+$
Decays\footnote{To be submitted to Phys.\ Rev.\ D.}}

\bigskip
\medskip

\normalsize
{\it Jonathan L. Rosner and Denis A. Suprun \\
\medskip

Enrico Fermi Institute and Department of Physics \\
University of Chicago, Chicago, Illinois 60637 } \\

\bigskip
\bigskip
{\bf ABSTRACT}

\end{center}

\begin{quote}
In a recently suggested method for measuring the weak phase $\gamma$
in $B^\pm \to K^\pm (KK^*)_D$ decays, the relative strong phase $\delta_D$ in
$D^0 \to K^{*+} K^-$ and $D^0 \to K^{*-} K^+$ decays (equivalently,
in $D^0 \to K^{*+} K^-$ and $\od \to K^{*+} K^-$) plays a role.  It is shown
how a study of the Dalitz plot in $D^0 \to K^+ K^- \pi^0$ can yield
information on this phase, and the size of the data sample which would give a
useful measurement is estimated.
\end{quote}

\leftline{PACS numbers: 13.25.Ft; 13.25.-k; 14.40.Lb}
\bigskip

The relative strong phases for charmed particle decays obey patterns which 
are not easily anticipated from first principles but are subject to detailed
experimental study, for example through the construction of amplitude
triangles based on experimentally observed decay rates 
\cite{Suz,JRFSa,JRFSb,JRChZu}.
It has also been suggested \cite{Falk,Berg,Lip} that the final-state
phase in the doubly Cabibbo-suppressed decay $D^0 \to K^+ \pi^-$ may not
be the same as that in the Cabibbo-favored decay $D^0 \to K^- \pi^+$,
even though they should be equal in the flavor-SU(3) limit \cite{Wolf}.
Methods for measuring their difference have been proposed \cite{Xing,GGY}.
A Dalitz-plot method for measuring the corresponding phase difference
in $D^0 \to K^{*+} \pi^-$ and $D^0 \to K^{*-} \pi^+$ makes use of the
interference between $K^{*+}$ and $K^{*-}$ bands in $D^0 \to K_S \pi^+ \pi^-$
and is compatible with zero strong phase difference \cite{CLEODal,Palano}.

Recently the question has been raised of the relative strong phase $\delta_D$
between $D^0 \to K^{*+} K^-$ and $D^0 \to K^{*-} K^+$ decays (equivalently,
in $D^0 \to K^{*+} K^-$ and $\od \to K^{*+} K^-$) \cite{GLS}.  This phase
is important in a proposed method for measuring the weak phase $\gamma$ in
the $B^\pm \to (K K^*)_D K^{\pm}$ decays.
In the present note we 
point out that $\delta_D$ may be measured very directly through the
interference of $K^{*+}$ and $K^{*-}$ bands in $D^0 \to K^+ K^- \pi^0$ decays
\cite{ADS}.  We discuss the size
of present and anticipated samples of this final state and indicate the
attainable experimental precision for $\delta_D$.

We follow the notations of Ref.~\cite{GLS} and define the $D$ 
decay amplitudes
\beq
A_D\equiv A(D^0 \to K^- K^{*+}), \ \ \ \bar{A}_D\equiv A(\od\to K^- K^{*+} ),
\eeq
and their ratio
\beq
\frac{\bar{A}_D}{A_D}=r_D e^{i\delta_D}.
\label{eq:rD}
\eeq
The weak phase of $\od\to K^- K^{*+} $ is negligible, so the CP conjugate 
amplitude is $A(D^0 \to K^+ K^{*-} )=\bar{A}_D$.
We further define
\beq
A'_D\equiv A(D^0 \to K^- \,(K^+ \pi^0)_{K^{*+}} ), \ \ \ 
\bar{A'}_D\equiv  A(D^0 \to K^+ \,(K^- \pi^0)_{K^{*-}} ).
\label{eq:rD'}
\eeq
The amplitudes of the $K^{*+} \to K^+ \pi^0$ and $K^{*-} \to K^- \pi^0$ 
decays are equal. Then the ratio of the amplitudes in~(\ref{eq:rD'}) is
\beq
\frac{\bar{A'}_D}{A'_D}=\frac{\bar{A}_D}{A_D}=r_D e^{i\delta_D}.
\eeq

\begin{figure}[p]
\centerline{\includegraphics[width=.65\textwidth]{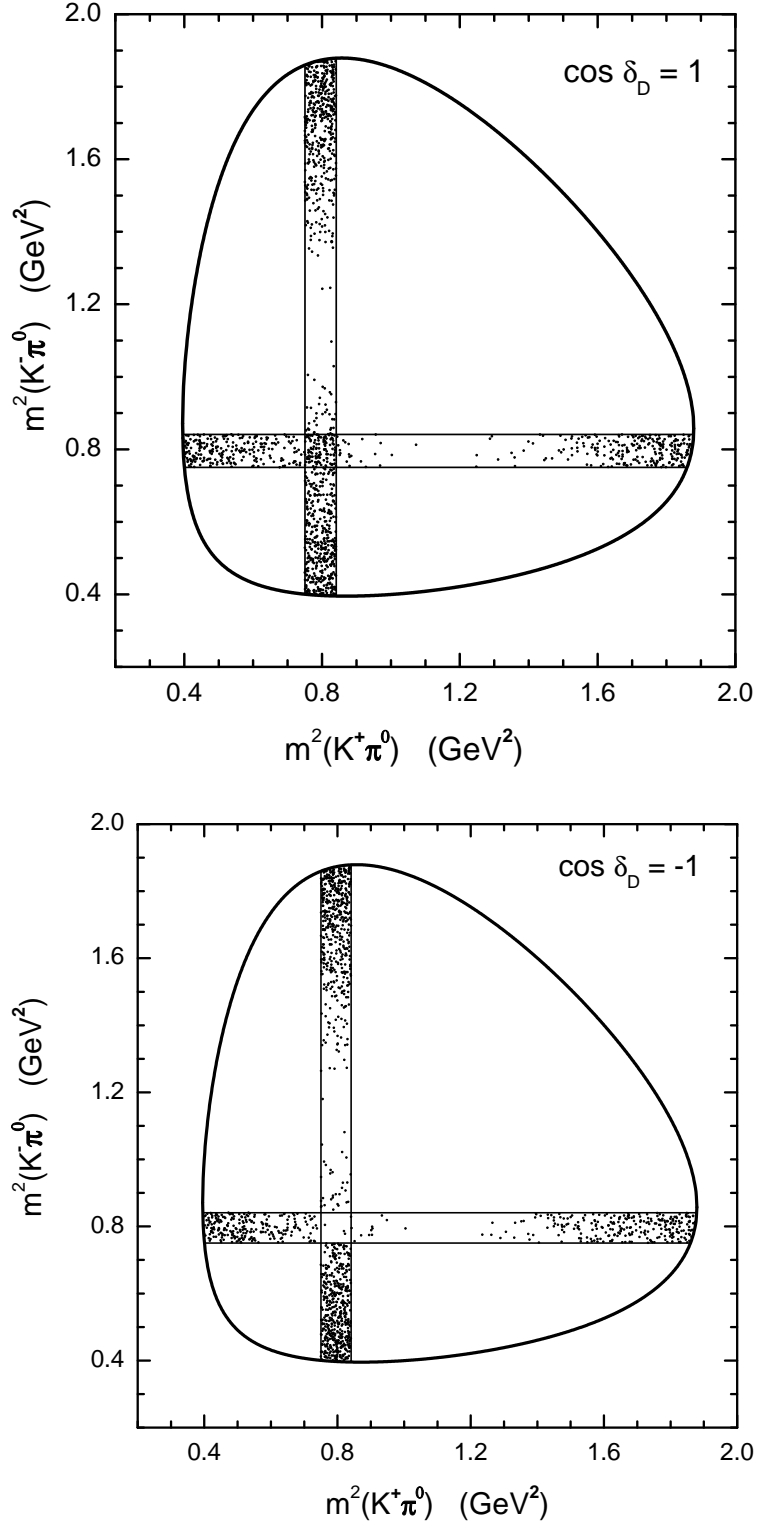}}
\caption{The Dalitz plots of the $D^0 \to K^+ K^- \pi^0$ decay. Top panel: 
constructive interference ($\cos\delta_D=1$), 113 events in the square 
region; bottom panel: destructive interference ($\cos\delta_D=-1$), 4 events 
in the square region. The total number of events in the bands is 
$N=1500$ in both cases. }
\label{fig:Dalitz}
\end{figure}

Two channels of $D^0 \to K^+ K^- \pi^0$ go through a resonant decay of an
intermediate $K^{*+}$ or $K^{*-}$. They fill two bands in the Dalitz plot
(see Fig.~\ref{fig:Dalitz}). The width of these bands is determined by the 
full width $\Gamma\equiv\Gamma_{K^{*\pm}}=(50.8\pm0.9)$~MeV~\cite{PDG}. Namely,
the left vertical line corresponds to 
$m^2_{K^+\pi^0}=(m_{K^{*+}}-\Gamma/2)^2$, while the right one corresponds to 
$m^2_{K^+\pi^0}=(m_{K^{*+}}+\Gamma/2)^2$. Analogous expressions determine the
values of $m^2_{K^-\pi^0}$ along the bottom and top borders of the horizontal
band.  For now we will neglect the actual Breit-Wigner distribution of event 
density across the bands. Instead, we will assume that the resonant decays 
are equally likely to appear near the central line of a band and near its 
borders. We will also assume that the resonant decays do not fall in the 
regions outside the two bands.  
We will neglect other resonant decays with
smaller branching ratios that are not yet detected but may contribute to the
Dalitz plot, such as 
$D^0 \to \pi^0 \,(K^+ K^-)_{\phi}$
$D^0 \to \pi^0 \,(K^+ K^-)_{a_0}$,
$D^0 \to \pi^0 \,(K^+ K^-)_{f_0}$, 
$D^0 \to K^- \,(K^+ \pi^0)_{K^*_0(1430)^+}$, and
$D^0 \to K^- \,(K^+ \pi^0)_{\kappa(800)^+}$.
Some of them are discussed later in the text and in Appendix~B. Non-resonant 
decays uniformly 
fill the allowed phase space and provide a small background. 
For simplicity of the argument we will neglect it as well. 

\begin{figure}[p]
\centerline{\includegraphics[width=.65\textwidth]{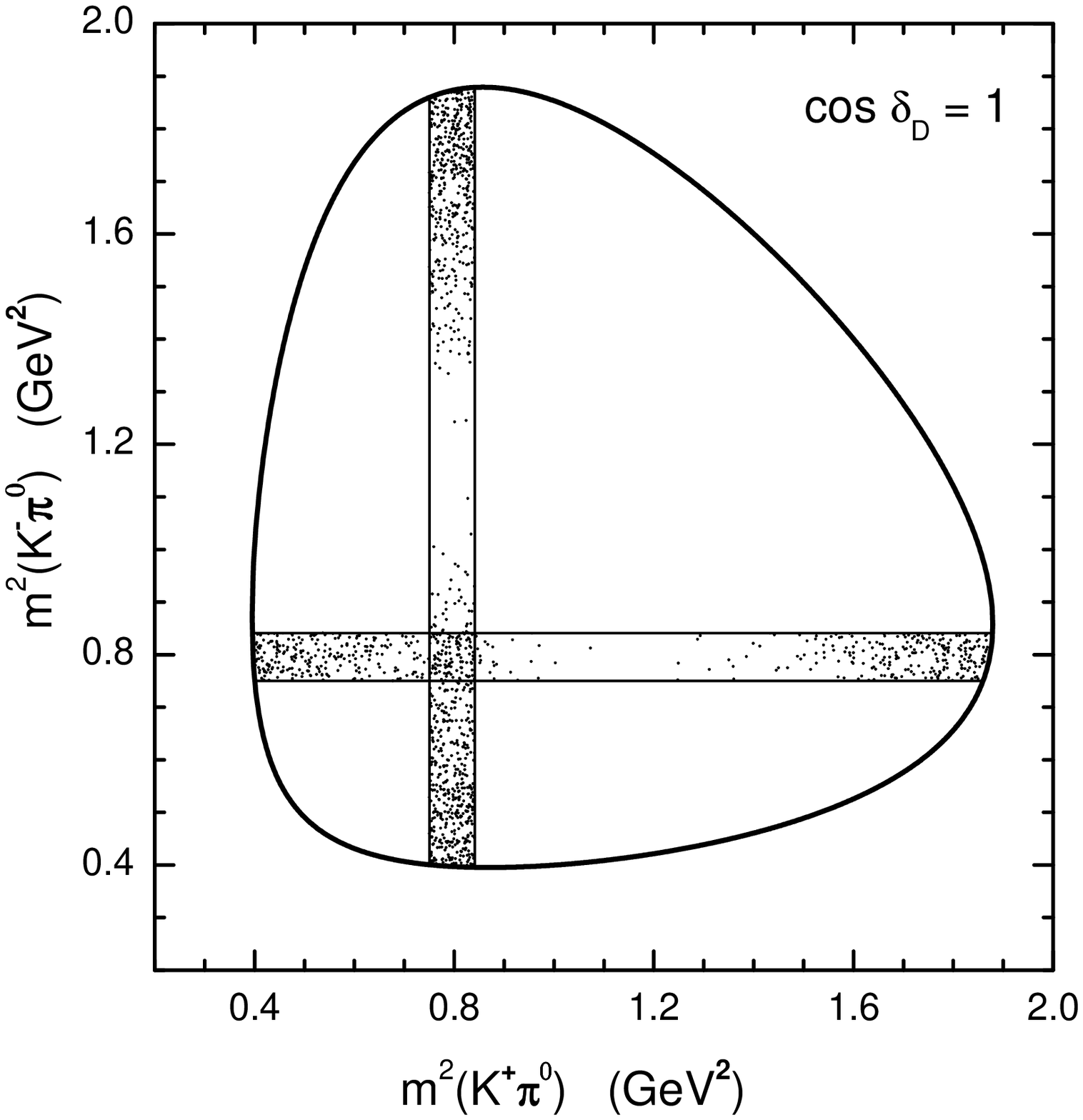}}
\end{figure}

The square at the intersection of the bands is the region where two channels 
interfere with each other. We denote $\epsilon$ to be the fraction of $D^0 
\to K^- \,(K^+ \pi^0)_{K^{*+}}$ decays that fall into the square region. 
This fraction only depends on masses and spins of particles involved in the 
process and the width $\Gamma=\Gamma_{K^{*\pm}}$. So, the probability of a 
$D^0 \to K^+ \,(K^- \pi^0)_{K^{*-}}$ decay falling into the square region is 
$\epsilon$ as well. This probability is calculated in Appendix~A: 
$\epsilon\approx0.039$. 

Now we can write the number of decays detected in the square region of the 
Dalitz diagram:
\beq
N_s \propto |\sqrt{\epsilon} A'_D+\sqrt{\epsilon}\bar{A'}_D|^2
= \epsilon\,(1+2r_D\cos\delta_D+r_D^2)\,|A'_D|^2 \ \ ,
\eeq
while the rest of the resonant decays contribute to the bands outside the 
square region:
\beq
N_{out}\propto (1-\epsilon)\,(|A'_D|^2+|\bar{A'}_D|^2)
= (1-\epsilon)\,(1+r_D^2)\,|A'_D|^2 \ \ ,
\eeq
so that the total number of the events detected in the bands is
\beq
N=N_s+N_{out} \propto (1+2\epsilon r_D\cos\delta_D+r_D^2)\,|A'_D|^2 \ \ .
\eeq

Experimental measurements of $N_s$ and $N$ provide a way of measuring the 
strong phase $\delta_D$:
\beq
\cos\delta_D=\frac{1+r_D^2}{2\epsilon r_D}\,\frac{N_s/N-\epsilon}{1-N_s/N} 
\ \ .
\label{eq:cos}
\eeq
The uncertainty in $\epsilon$ can be neglected because it is determined by 
the uncertainties in particles' masses and width $\Gamma$, which are small.
The ratio $r_D$ defined by Eq.~(\ref{eq:rD}) can be calculated from the 
measured branching ratios: ${\cal B}(D^0 \to K^+ K^{*-})=(2.0\pm1.1) \cdot 
10^{-3}$ and ${\cal B}(D^0 \to K^- K^{*+})=(3.8\pm0.8) \cdot 
10^{-3}$~\cite{PDG}. Assuming the uncertainties of these two measurements 
are uncorrelated, $r_D=0.73\pm0.21$. 
These values are based on a sample of 35~$D^0 \to K K^*$ 
decays~\cite{CLEO_ratios}. For a larger sample, the relative 
uncertainty in $r_D$ will decrease as $1/\sqrt{N}$.
Taking the uncertainties of the decay numbers $N_s$ and $N$ to be their 
square roots, we can calculate the uncertainty $\sigma(\cos\delta_D)$.
One can show that the uncertainty in $\cos\delta_D$ is mostly determined by 
the uncertainty in $N_s$:
\beq
\sigma(\cos\delta_D)\approx\left|\frac{\partial \cos\delta_D}
{\partial N_s}\right|\,\sigma(N_s)=
\frac{(1-\epsilon)\,(1+r_D^2)}{2 \epsilon r_D}\, 
\frac{\sqrt{N_s/N}}{(1-N_s/N)^2}\,\frac{1}{\sqrt{N}} \ \ ,
\label{eq:sigcos}
\eeq
Unlike $\cos\delta_D$ itself, the uncertainty of this quantity depends not 
only on the ratio $N_s/N$ but on the total number $N$ of the events detected 
in the bands as well. 

As an aside, note that Eq.~(\ref{eq:cos}) predicts a linear dependence of 
$\cos\delta_D$ 
on $Z \equiv (N_s/N)/(1-N_s/N)$ with the slope $S=(1-\epsilon)(1+r_D^2)/ 
(2\epsilon r_D)$. We could alternatively write Eq.~(\ref{eq:sigcos}) as
\beq
\sigma(\cos\delta_D)\approx\left|\frac{\partial \cos\delta_D}
{\partial Z}\right|\,\sigma(Z)\approx S\,
\frac{\sqrt{N_s/N}}{(1-N_s/N)^2}\,\frac{1}{\sqrt{N}} \ \ .
\label{eq:sigZ}
\eeq

The maximum possible value of the ratio $N_s/N$ is achieved if the 
contributions from two bands are fully coherent, i.e., if $\cos\delta_D=1$. 
In this case
\beq
\frac{N_s}{N}=\left(\frac{N_s}{N}\right)_{max}= \frac{\epsilon(1+r_D)^2}
{1+2\epsilon r_D+r_D^2}=0.074\pm0.003 \ \ .
\eeq 
The minimum possible $N_s/N$ is a result of the fully destructive 
interference at $\cos\delta_D=-1$. Then,
\beq
\frac{N_s}{N}=\left(\frac{N_s}{N}\right)_{min}= \frac{\epsilon(1-r_D)^2}
{1-2\epsilon r_D+r_D^2}=0.0020\pm0.0035 \ \ .
\eeq 
Thus, if $\cos\delta_D$ is close to $-1$, one may observe no events in 
the square region. 
The source of the uncertainties in the maximum and minimum values of the 
$N_s/N$ ratio is the current 30\% error in $r_D$ which will be improved as 
more $D^0 \to K K^*$ decays are detected.  Within 1$\sigma$ uncertainty, we can
expect the $N_s/N$ ratio to lie between $0$ and $0.077$.

\begin{figure}[t]
\centerline{\includegraphics[width=.85\textwidth]{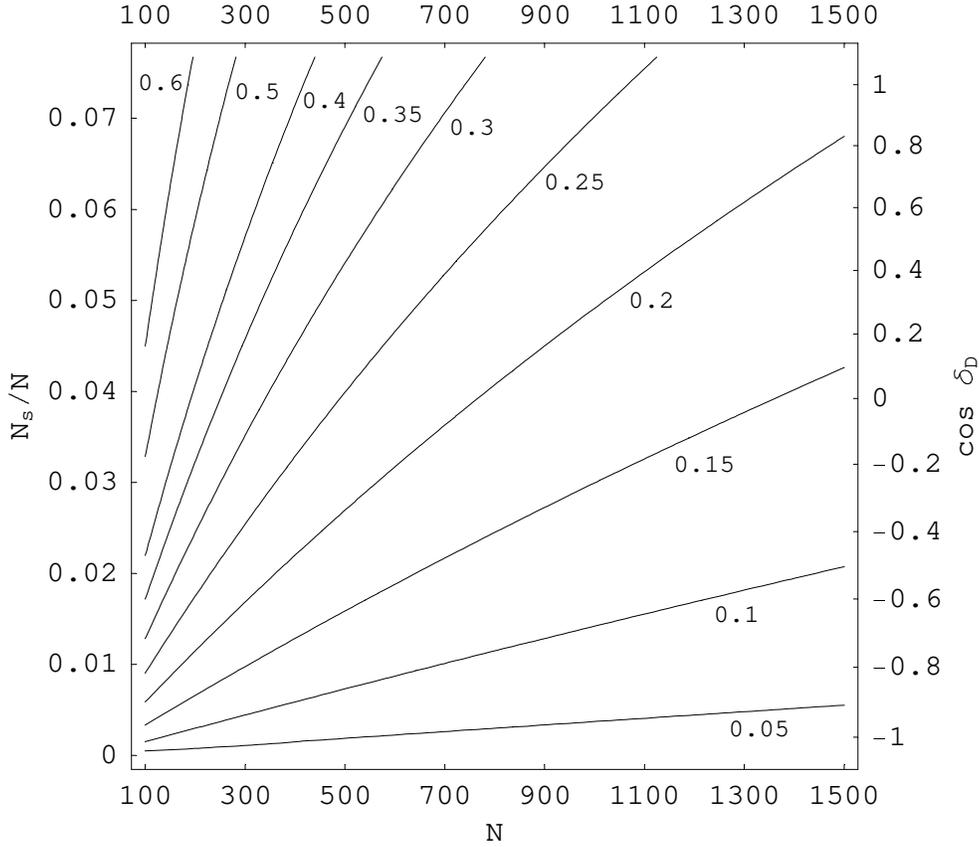}}
\caption{Contours of $\sigma(\cos\delta_D)$ for $N_s/N$ between $0$ and 
$0.077$, i.e., for $\cos\delta_D$ between $-1.05$ and $1.09$.}
\label{fig:contours}
\end{figure}

Figure~\ref{fig:contours} shows the contours of constant 
$\sigma(\cos\delta_D)$ calculated for this region of $N_s/N$ from 
Eq.~(\ref{eq:sigcos}) for the total number of band events $N$ between 100
and 1500. The uncertainty in $\cos\delta_D$ is an increasing function 
of $N_s/N$. So, $\cos\delta_D$ will be measured 
least precisely if it is close to unity. This corresponds to a near maximum 
value of the $N_s/N$ ratio. To estimate the largest uncertainty for 
different numbers of band events, we calculate how $\sigma(\cos\delta_D)$ 
decreases with $N$ when $N_s/N$ is fixed at its maximum value of $0.077$:
\beq
\sigma_{max}(\cos\delta_D)
\approx \frac{8.4}{\sqrt{N}} \ \ .
\eeq

\begin{figure}[p]
\centerline{\includegraphics[width=.65\textwidth]{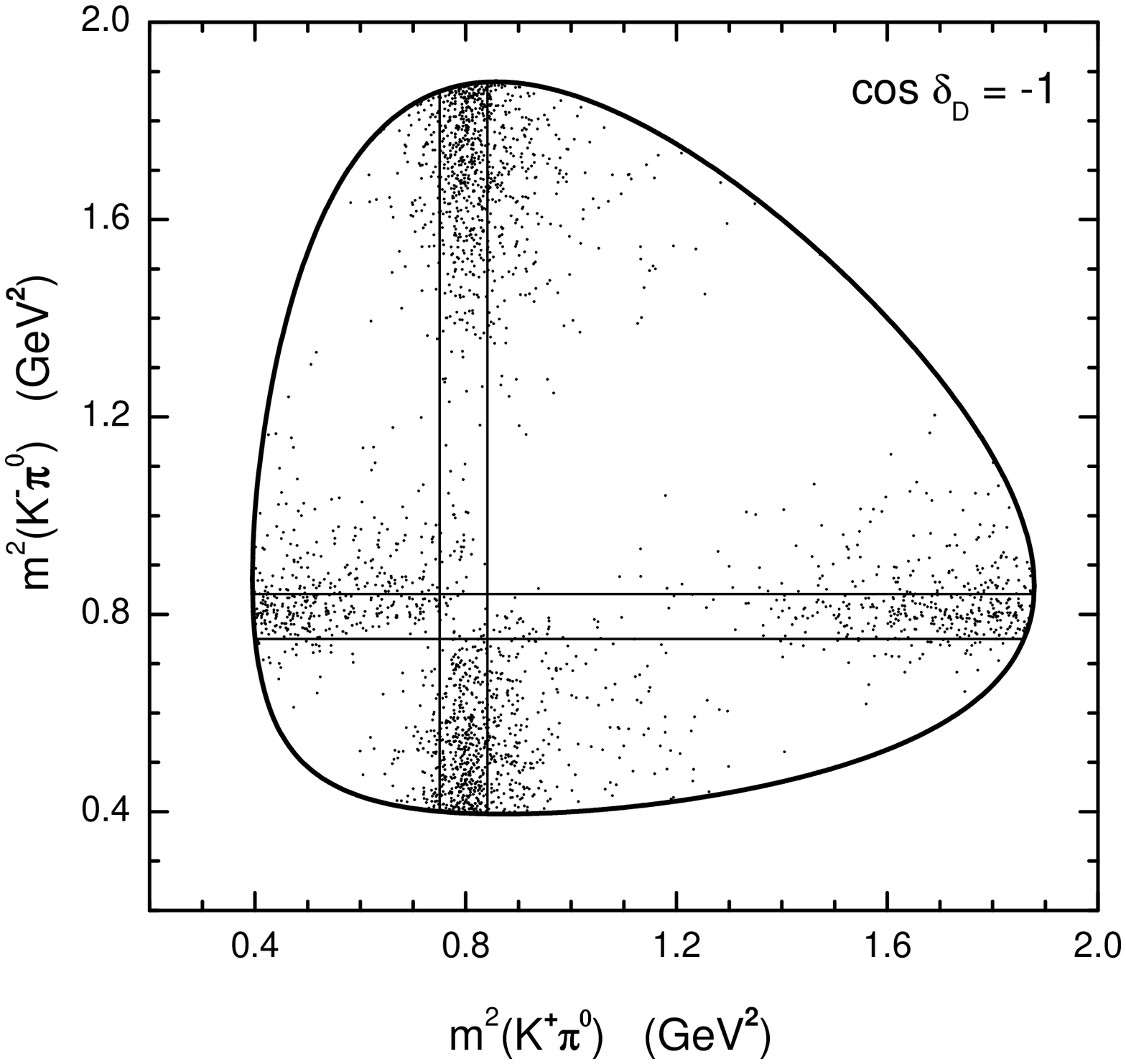}}
\caption{Two examples of realistic Dalitz plots of the 
$D^0 \to K^+ K^- \pi^0$ decay. 
Top panel: 
constructive interference ($\cos\delta_D=1$), 88 events in the square 
region; bottom panel: destructive interference ($\cos\delta_D=-1$), 18 
events in the square region. The total number of events in the bands is 
$N=1500$ in both cases. }
\label{fig:dist}
\end{figure}

Now we discuss the consequences of the fact that the event density across a 
resonant decay band is not uniform but follows the Breit-Wigner distribution.
The differential cross-section for any point on the Dalitz plot (see
Appendix~A) is
\beq
\frac{d^2\,\Gamma}{dm^2_{K^+\pi^0}\,dm^2_{K^-\pi^0}} \propto 
\left|\frac{A_1(m_{K^+\pi^0},\,m_{K^-\pi^0})} 
{m^2_{K^+\pi^0}-m^2_{K^{*+}}+i\,m_{K^{*+}}\Gamma} +
\frac{r_D e^{i\delta_D}\,A_2(m_{K^+\pi^0},\,m_{K^-\pi^0})} 
{m^2_{K^-\pi^0}-m^2_{K^{*-}}+i\,m_{K^{*-}}\Gamma} \right|^2  \ \,
\label{eq:dist}
\eeq
The Breit-Wigner factors in the denominators make the population density 
nonuniform across the bands while the kinematic factors 
$A_1(m_{K^+\pi^0},\,m_{K^-\pi^0})$ and \\
$A_2(m_{K^+\pi^0},\,m_{K^-\pi^0})$ 
are responsible for a characteristic emptiness in the middle of the bands.
The results of a Monte Carlo simulation of the 
distribution~(\ref{eq:dist}) are shown in Fig.~\ref{fig:dist}.

\begin{figure}[p]
\centerline{\includegraphics[width=.65\textwidth]{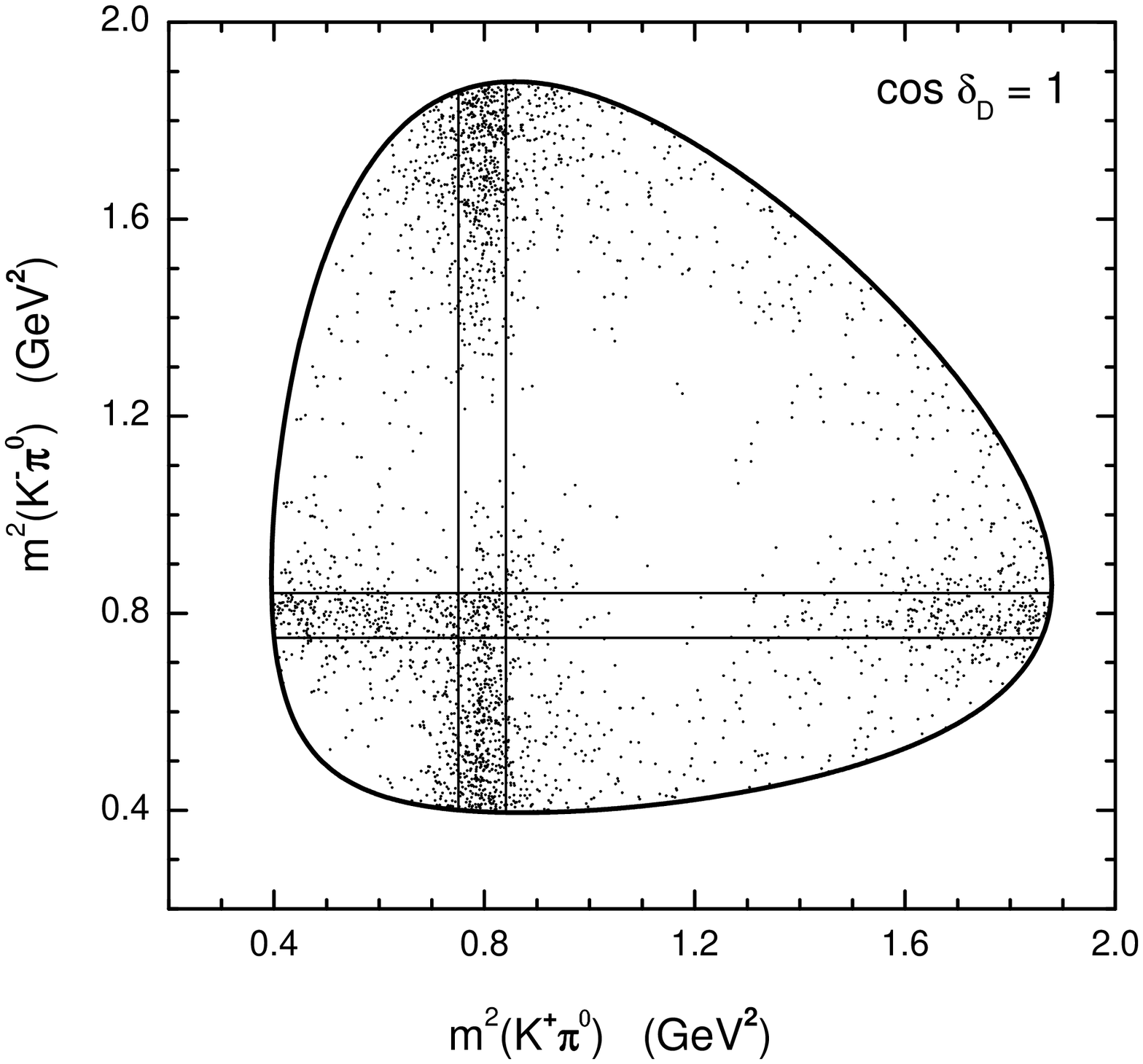}}
\end{figure}

\begin{figure}[t]
\centerline{\includegraphics[width=.85\textwidth]{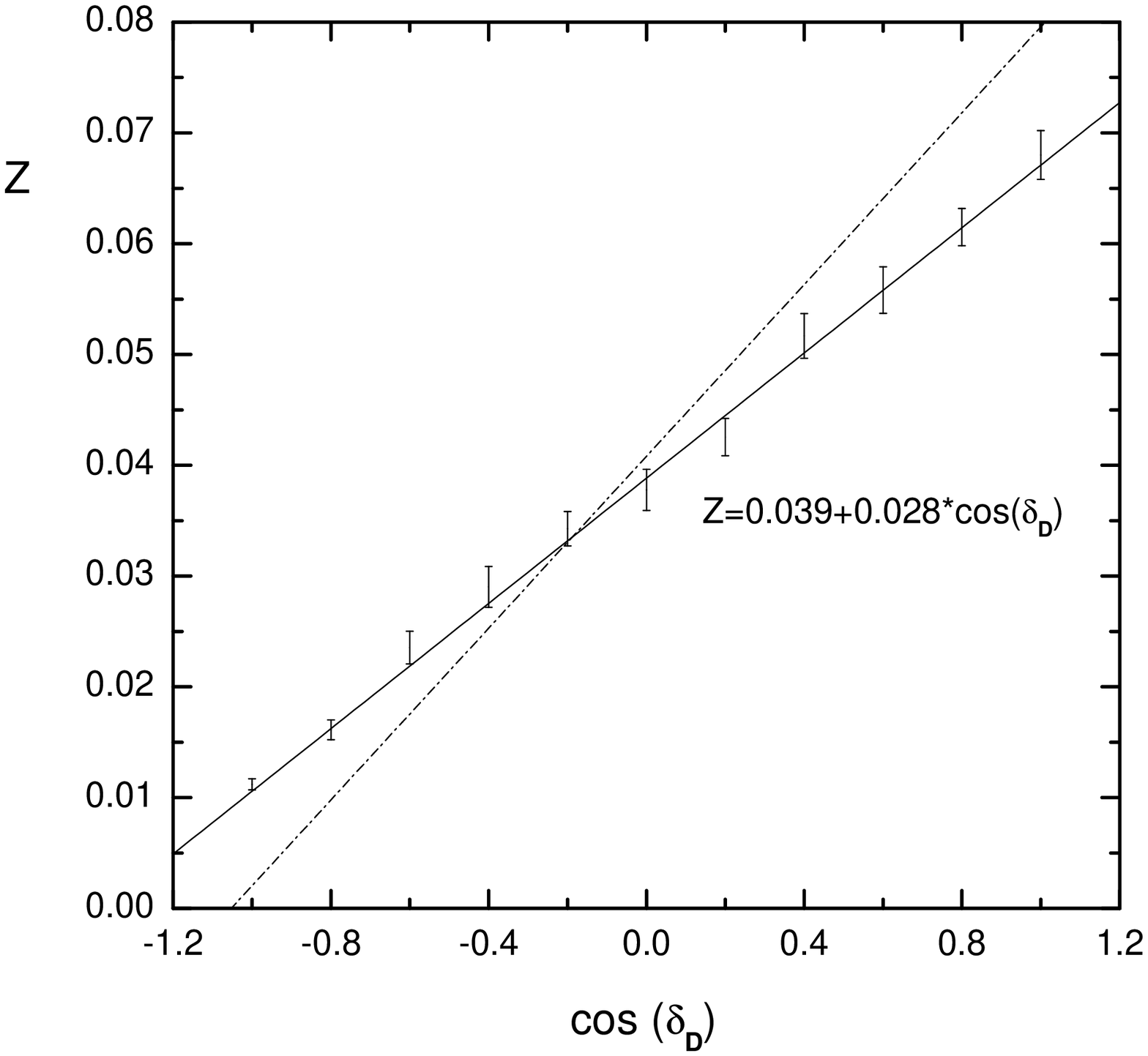}}
\caption{$Z \equiv (N_s/N)/(1-N_s/N)$ as a function of $\cos\delta_D$.
The solid line with the slope of $0.0283\pm0.0005$ is the best linear 
fit to the results of the Monte Carlo 
simulations. The dash-dotted line is the prediction of the simplified 
model which doesn't take into account the Breit-Wigner resonant shapes 
(Eq.~(\ref{eq:cos})).}
\label{fig:Z}
\end{figure}

We simulated the Dalitz plot distributions 10 times for each of 11 values of 
$\cos\delta_D$ between $-1$ and $1$.  For the purposes of these simulations we
assumed that $r_D$ is equal to its current central value of 0.73.
The plot of $Z \equiv (N_s/N)/(1-N_s/N)$
as a function of $\cos\delta_D$ is shown in Fig.~\ref{fig:Z}.  To estimate 
$\sigma(\cos\delta_D)$ we will
assume that the linear relationship between the two quantities still holds.
Then, the slope is $S=1/(0.0283\pm0.0005)=35.3\pm0.6$ while the maximum 
value of $N_s/N$ is 
$0.0637\pm0.0019$ at $\cos\delta_D=1$. 
Both errors are purely statistical Monte Carlo uncertainties.
These new values of the slope $S$ and $(N_s/N)_{max}$
can be plugged into Eq.~(\ref{eq:sigZ}) to give 
our best estimate of the maximum uncertainty in $\cos\delta_D$:
$\sigma(\cos\delta_D)=(10.16\pm0.26)/\sqrt{N}$, with the 
upper bound
\beq
\sigma_{max}(\cos\delta_D)
\approx \frac{10.4}{\sqrt{N}} \ \ .
\eeq

Thus, we see that the most precise measurements will be made if 
$\cos\delta_D$ is close to $-1$. The 
uncertainty of the least precise measurements (in case $\cos\delta_D$ is 
unity) becomes smaller than 0.33 at $N\approx1000$. Although this uncertainty
is rather large, it at least allows one to distinguish $\cos\delta_D$ from 0.
The measurement of $\cos\delta_D$ will be improved to reach the uncertainty 
of 0.27 or better when 1500 resonant events are detected in the bands.

In fact, 1500 resonant decays in the bands is the largest sample one can 
expect from CLEO-c. The CESR accelerator will operate at a center-of-mass 
energy of $\sqrt{s}\sim3.77$~GeV ($\psi''$) for approximately one year. The 
anticipated integrated luminosity will reach $3$~fb$^{-1}$.  This corresponds
to a sample of 30~million $D\bar{D}$ pairs, with 17.5~million of them being
$D^0 \od$ pairs. The expected sample will exceed the Mark~III experiment 
dataset by a factor of 300. Approximately 5~million of $D^0$ and $\od$ 
mesons will be flavor tagged~\cite{CLEOc}. The other $D$ of a pair may 
decay to the $K^+K^-\pi^0$ final state through an intermediate $K^*$. 
The branching 
ratios of these resonant decays are 
${\cal B}(D^0 \to K^+ \,(K^- \pi^0)_{K^{*-}})= 
\frac13\,(2.0\pm1.1) \cdot 10^{-3}$ and 
${\cal B}(D^0 \to K^- \,(K^+ \pi^0)_{K^{*+}} )= 
\frac13\,(3.8\pm0.8) \cdot 10^{-3}$, adding up to about $2\cdot 10^{-3}$. 
Neglect interference effects and the number of decays should be 
around 10000. The estimated reconstruction efficiency for these 3-body 
decays is approximately 30\%, so 3000 events will be detected. 
The Breit-Wigner distribution dictates that the bands of the Dalitz plot 
will be populated by half of these, i.e., by 1500~events. 

The method that will be used in data analysis will likely adopt the 
multi-variable fitting described in~\cite{Kopp} and~\cite{E791 sigma} 
instead of taking a close look at the number of events in the square region. 
We hope, however, that 
this note gives a good estimate of the expected uncertainty and its 
dependence on the total number of detected $D^0 \to K^+ K^{*-}$ and 
$D^0 \to K^- K^{*+}$ events.
Other resonant decays with smaller branching 
ratios, 
$D^0 \to \pi^0 \,(K^+ K^-)_{\phi}$
$D^0 \to \pi^0 \,(K^+ K^-)_{a_0}$,
$D^0 \to \pi^0 \,(K^+ K^-)_{f_0}$, 
$D^0 \to K^- \,(K^+ \pi^0)_{K^*_0(1430)^+}$, and
$D^0 \to K^- \,(K^+ \pi^0)_{\kappa(800)^+}$, 
may contribute to the Dalitz plot. The estimate of the uncertainty is most
sensitive to the number of events inside the square region.
Unless the bands of those decays overlap with it, they should not
considerably change our estimate.

Among the five decays listed above, only those of the $\kappa(800)^+$
have the potential to 
contribute to the
square region. However, the $\kappa$ is not likely to be among the 
intermediate states that make a significant contribution to 
$D^0 \to K^- K^+ \pi^0$ decays (see Appendix~B).
The $\phi$ meson is a narrow vector resonance which is not much
heavier than the combined mass of two charged $K$ mesons. 
Therefore, it could 
only produce a narrow diagonal band at the very edge of the Dalitz plot. 
Its presence would not change the $K^*$ band population. The same is true 
for $a_0(980)$ and $f_0(980)$ decays. 
They are lighter and broader ($40-100$~MeV) but yet not broad enough to 
significantly affect even the outer ends of the $K^*$ bands. 
Such a possibility is present for $K^*_0(1430)$ decays. 
The square region lies outside the $K^*_0(1430)$
bands and their impact on the number of events $N_s$ inside the square is 
insignificant. They can only make a relatively small contribution to the 
total number of band events $N$ which would add just a small correction to 
the uncertainty in the strong phase $\delta_D$.

\section*{Acknowledgments}

We wish to thank D. M. Asner and M. Gronau for helpful correspondence.
This work was supported in part by the United States Department of
Energy through Grant No.\ DE FG02 90ER40560.  

\section*{Appendix~A: Kinematics and decay amplitudes}

\begin{figure}[t]
\centerline{\includegraphics[width=.5\textwidth]{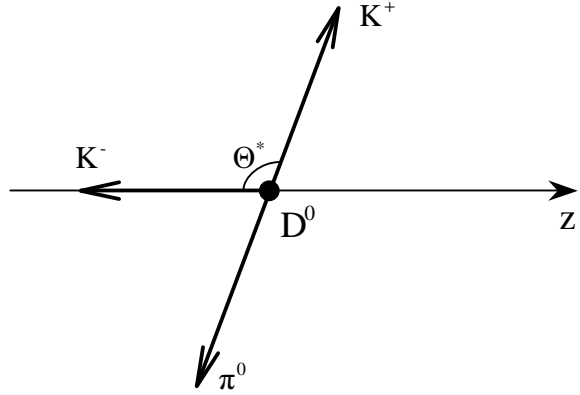}}
\caption{The  $D^0 \to K^- \,(K^+ \pi^0)_{K^{*+}}$ decay in
the rest frame of $K^+$ and $\pi^0$.}
\label{fig:decay}
\end{figure}

The first stage of the $D^0 \to K^- \,(K^+ \pi^0)_{K^{*+}}$ process is the
decay of a pseudoscalar meson $D^0$ into a pseudoscalar $K^-$ and a 
(possibly off-shell) vector $K^{*+}$.  Afterwards, the latter decays into $K^+$
and $\pi^0$.  From angular momentum conservation the helicity of $K^{*+}$ is
$0$.  The corresponding polarization vector is $\epsilon_{K^{*+}} =
\epsilon^{(\lambda=0)} = (|{\bf p}_{K^{*+}}|, 0, 0, E_{K^{*+}})/m_{K^+\pi^0}$.
($m_{K^+\pi^0}$ is the invariant mass of $K^{*+}$ and the $z$ axis is chosen 
to point in the direction of the $K^{*+}$ momentum ${\bf p}_{K^{*+}}$,
see Fig.~\ref{fig:decay}). 

The amplitude $A_1(m_{K^+\pi^0},\,m_{K^-\pi^0})$ of the 
$K^{*+}\to K^+ \pi^0$ decay should be Lorentz
invariant, i.e., it should contain a product of two 4-vectors.
There is only one non-vanishing possibility, 
$\epsilon_{K^{*+}}\,(p_{K^+}-
p_{\pi^0})$, since the other, $\epsilon_{K^{*+}}\,(p_{K^+}+p_{\pi^0})
=\epsilon_{K^{*+}}\,p_{K^{*+}}$, is identically zero.
Then the former can be written in the rest frame of $K^+$ and $\pi^0$ as 
\beq
A_1(m_{K^+\pi^0},\,m_{K^-\pi^0}) \propto
(0, 0, 0, 1)\,
(E_{K^+}^*-E_{\pi^0}^*, 2\,{\bf p}_{K^+}^*)=2\,|{\bf p}_{K^+}^*|\,
\cos\theta^* \ \ , 
\eeq
where $\theta^*$ is the angle between the negative direction of
the $z$ axis and the direction of the $K^+$ momentum ${\bf p}_{K^+}^*$ in 
the rest frame of $K^+$ and $\pi^0$. We will keep
using the ``*" subscript for quantities determined in this frame.
%
$\cos\theta^*$ is given by
\beq
\cos\theta^*=\frac{m^2_{K^-\pi^0}- m^2_{K^-}-m^2_{\pi^0}-2E_{K^-}^*E_{\pi^0}^*}
{2|{\bf p}_{K^-}^*||{\bf p}_{\pi^0}^*|} \ \ ,
\label{eq:coscm}
\eeq
so 
\beq
A_1(m_{K^+\pi^0},\,m_{K^-\pi^0}) \propto 
\frac{m^2_{K^-\pi^0}- m^2_{K^-}-m^2_{\pi^0}-2E_{K^-}^*E_{\pi^0}^*}
{|{\bf p}_{K^-}^*|} \ \ ,
\eeq
where 
\beq
E_{K^-}^*=(m^2_{D^0}-m^2_{K^-}-m^2_{K^+\pi^0})/2m_{K^+\pi^0} \ \ ,
\eeq
\beq 
E_{\pi^0}^*=(m^2_{K^+\pi^0}-m^2_{K^+}+m^2_{\pi^0})/2m_{K^+\pi^0} \ \ , 
\eeq
\beq
|{\bf p}_{K^-}^*|=\lambda^{1/2}(m^2_{D^0}, m^2_{K^-}, m^2_{K^+\pi^0})/
2m_{K^+\pi^0} \ \ ,
\eeq
\beq
|{\bf p}_{\pi^0}^*|=|{\bf p}_{K^+}^*| \ \ ,
\eeq
\beq
\lambda(x, y, z)\equiv x^2+y^2+z^2-2xy-2xz-2yz \ \ .
\eeq
Including the finite resonance width $\Gamma_{K^{*+}}$ into the $K^{*+}$
propagator, we can
write the amplitude of the $D^0 \to K^- \,(K^+ \pi^0)_{K^{*+}}$ decay as
\beq
A(D^0 \to K^- \,(K^+ \pi^0)_{K^{*+}}) \propto 
\frac{A_1(m_{K^+\pi^0},\,m_{K^-\pi^0})} 
{{m^2_{K^+\pi^0}-m^2_{K^{*+}}+i\,m_{K^{*+}}\Gamma}} \ \ . 
\label{eq:ampl1}
\eeq
As for the $D^0 \to K^+ \,(K^- \pi^0)_{K^{*-}}$ decay, its amplitude can 
be derived in a similar way and is equal to 
\beq
A(D^0 \to K^+ \,(K^- \pi^0)_{K^{*-}}) \propto 
r_D e^{i\delta_D}\,\frac{A_2(m_{K^+\pi^0},\,m_{K^-\pi^0})} 
{m^2_{K^-\pi^0}-m^2_{K^{*-}}+i\,m_{K^{*-}}\Gamma} \ \ ,
\label{eq:ampl2}
\eeq
with the kinematic factor $A_2$ defined as 
$A_2(m_{K^+\pi^0},\,m_{K^-\pi^0})\equiv A_1(m_{K^-\pi^0},\,m_{K^+\pi^0})$.
The factor $r_D e^{i\delta_D}$ accounts for possible differences in
hadronization as vector particles between quarks arising from the virtual 
$W^+$
and spectator quarks.

\subsection*{Calculation of the fraction $\epsilon$ of resonant decays that 
fall into the square region}

For the particular case of an on-shell resonant $K^{*+}$ we can neglect 
the Breit-Wigner denominator of Eq.~(\ref{eq:ampl1}). In this case 
the amplitude of the
$D^0 \to K^- \,(K^+ \pi^0)_{K^{*+}}$ decay is proportional to
$A_1(m_{K^{*+}},\,m_{K^-\pi^0})$.
The kinematics of the two-body $D^0 \to K^- K^{*+}$ and 
$K^{*+} \to K^+ \pi^0$ decays
determine $E_{K^-}^*=1.37$~GeV, $E_{\pi^0}^*=0.32$~GeV,
$|{\bf p}_{K^-}^*|=1.27$~GeV and $|{\bf p}_{\pi^0}^*|=0.29$~GeV. 
As a result, Eq.~(\ref{eq:coscm}) says:
\beq
\cos\theta^* = 1.36\,(m^2_{K^-\pi^0}-1.135) \ \ ,
\label{eq:bound1}
\eeq
where $m^2(K^- \pi^0)$ is in GeV$^2$. Thus, the amplitude of 
the $D^0 \to K^- \,
(K^+ \pi^0)_{K^{*+}}$ decays is proportional to $(m^2_{K^-\pi^0}-1.135)$.
These resonant decays fill the vertical band in a nonuniform way: no decays
happen at the middle of the band where $m^2_{K^-\pi^0}-1.135=0$. The majority
of the events will concentrate near both band ends where $|m^2_{K^-\pi^0}-
1.135|$ is the largest.

Now we can calculate the fraction of $D^0 \to K^- \,(K^+ \pi^0)_{K^{*+}}$ 
decays that fall into the square region,
\beq
\epsilon=\left. \int\limits_{0.75}^{0.84} (x-1.135)^2\,dx \right/
\int\limits_{0.40}^{1.87} (x-1.135)^2\,dx=0.039 \ \ ,
\eeq
where $(m_{K^{*+}}-\Gamma/2)^2=0.75$~GeV$^2$ and $(m_{K^{*+}}+\Gamma/2)^2=
0.84$~GeV$^2$ are the boundaries of the square region and $0.40$ and 
$1.87$ are the boundaries of the whole band. The latter can be derived from
Eq.~(\ref{eq:bound1}).

This simple calculation implied that the population density of the vertical
band is constant along any cross section of the band, i.e., at a fixed 
$m^2_{K^-\pi^0}$ it is independent of variations of $m^2_{K^+\pi^0}$ across 
the band. A more precise discussion involves a simulation of the 
interference between the Breit-Wigner resonant shapes of 
Eqs.~(\ref{eq:ampl1}) and (\ref{eq:ampl2}).

\section*{Appendix~B: Influence of scalar resonance $\kappa$}

The existence of broad scalar resonances below $1$~GeV has been a 
controversial issue for a long time~\cite{theory}. A few experiments 
have been able
to explore the possibility of their presence as intermediate resonant 
states in three-body $D$ decays. The modes that were studied include 
$D^+ \to \pi^- \pi^+ \pi^+$~\cite{E791 sigma}, 
$D^+_s \to \pi^- \pi^+ \pi^+$~\cite{E791 f0},
$D^+ \to K^- \pi^+ \pi^+$~\cite{E791 kappa} \ (E791 collaboration),
$D^0 \to K^0_S \pi^+ \pi^-$~\cite{CLEODal},
$D^0 \to K^- \pi^+ \pi^0$~\cite{Kopp} \ (CLEO), and 
$D^0 \to K^0 K^- \pi^+$~\cite{Babar} \ (BaBar).
The first two studies obtained evidence for a light 
($478$~MeV) $\sigma$ resonance and measured the properties of the 
$f_0(980)$. The last four might provide some information on the presence of
an intermediate S-wave $K\pi$ resonance. Indeed, the E791 analysis of a Dalitz
plot found that the best fit to the data is obtained allowing for the
presence of an additional scalar resonance $\kappa(800)^0$. However, neither
CLEO studies found evidence for $\kappa^0$ or its isodoublet partner 
$\kappa^+$. The preliminary BaBar analysis saw $\kappa$ at the level of
$1\sigma$ which does not allow the confirmation of its presence.
Other types of decays could also provide a glimpse of $\kappa$.  The
BES collaboration
found $\kappa^0$ as an intermediate state in $J/\psi \to \bar{K}^*(892)^0 
K^+ \pi^-$ decays~\cite{BES}, while the FOCUS collaboration studied the 
interference phenomena in $D^+ \to K^- \pi^+ \mu^+ \nu$ decays~\cite{FOCUS}. 
Their data can be described by $\bar{K}^{*0}$ interference with either a 
constant amplitude or a broad spin zero resonance.

The $D^0 \to K^\mp K^{*\pm}$ decays discussed in this note can be affected 
by the possible presence of $\kappa^\pm$ among the intermediate states. The 
bands of a broad $\kappa(800)$ would cover more than 50\% of the 
Dalitz plot, thereby interfering with the $K^*$ bands and affecting their 
population.  One would expect that in this case the total branching ratio of 
$D^0 \to K^+ K^- \pi^0$ decays would be considerably larger than the sum of 
the $D^0 \to KK^*$ modes. Indeed, in 
$D^+ \to K^- \pi^+ \pi^+$ an unusually high fraction (over 90\%) of decays 
was found to be non-resonant by previous experiments~\cite{previous}. That 
was unusual as the non-resonant (NR) contribution in three-body decays is small
in most other cases. That was an indication of a possible broad scalar
contribution and motivated the recent searches for it. It was found that the
complex structure of the Dalitz plot was best explained when the $\kappa$
presence is assumed \cite{E791 kappa}.
Then, intermediate decays through the $\kappa \pi^+$ state account for about
50\% of decays while the NR fraction drops to a value of 13\% more
characteristic of other decays.

The present knowledge of $D^0 \to K^\mp K^{*\pm}$ decays does not reveal a
similar large non-resonant (or broad scalar) contribution. The current data 
on the resonant~\cite{PDG,CLEO_ratios} and inclusive~\cite{PDG,inclusive} 
decays comes from CLEO measurements. The inclusive branching ratio is 
${\cal B}(D^0 \to K^+ K^- \pi^0)=(1.24\pm0.35) \cdot 10^{-3}$. The branching 
ratios of the $K\pi$ resonant decays are
${\cal B}(D^0 \to K^+ \,(K^- \pi^0)_{K^{*-}})= 
\frac13\,(2.0\pm1.1) \cdot 10^{-3}=(0.67\pm0.37) \cdot 10^{-3}$ and 
${\cal B}(D^0 \to K^- \,(K^+ \pi^0)_{K^{*+}} )= 
\frac13\,(3.8\pm0.8) \cdot 10^{-3}=(1.27\pm0.27) \cdot 10^{-3}$.
Neglecting the interference between these two channels (it affects just
about 4\% of these decays; see Appendix A), the two branching ratios add up 
to $(1.93\pm0.45) \cdot 10^{-3}$, consistent with the inclusive branching 
ratio within the current large uncertainties. Basically, there is no room 
for a broad scalar resonance 
channel. For example, it cannot negatively interfere with both halves of a 
$K^*$ band. The phase variation across it would be significant 
($\approx90^\circ$) for a $K^*$ channel and much smaller for a broad
$\kappa$ one. If this channel is strong enough to cancel half the $K^*$
decays it would contribute many times more than that outside the $K^*$ bands.
That would contradict the smallness of the inclusive branching ratio.
Thus, we conclude that a broad scalar $\kappa$, if present, 
could only comprise a small fraction of 
$D^0 \to K^+ K^- \pi^0$ decays and would not significantly affect the 
estimate of the uncertainty in the strong phase $\delta_D$.

\def \ajp#1#2#3{Am.\ J. Phys.\ {\bf#1}, #2 (#3)}
\def \apny#1#2#3{Ann.\ Phys.\ (N.Y.) {\bf#1}, #2 (#3)}
\def \app#1#2#3{Acta Phys.\ Polonica {\bf#1}, #2 (#3)}
\def \arnps#1#2#3{Ann.\ Rev.\ Nucl.\ Part.\ Sci.\ {\bf#1}, #2 (#3)}
\def \art{and references therein}
\def \cmts#1#2#3{Comments on Nucl.\ Part.\ Phys.\ {\bf#1}, #2 (#3)}
\def \cn{Collaboration}
\def \cp89{{\it CP Violation,} edited by C. Jarlskog (World Scientific,
Singapore, 1989)}
\def \efi{Enrico Fermi Institute Report No.\ }
\def \epjc#1#2#3{Eur.\ Phys.\ J. C {\bf#1}, #2 (#3)}
\def \f79{{\it Proceedings of the 1979 International Symposium on Lepton and
Photon Interactions at High Energies,} Fermilab, August 23-29, 1979, ed. by
T. B. W. Kirk and H. D. I. Abarbanel (Fermi National Accelerator Laboratory,
Batavia, IL, 1979}
\def \hb87{{\it Proceeding of the 1987 International Symposium on Lepton and
Photon Interactions at High Energies,} Hamburg, 1987, ed. by W. Bartel
and R. R\"uckl (Nucl.\ Phys.\ B, Proc.\ Suppl., vol.\ 3) (North-Holland,
Amsterdam, 1988)}
\def \ib{{\it ibid.}~}
\def \ibj#1#2#3{~{\bf#1}, #2 (#3)}
\def \ichep72{{\it Proceedings of the XVI International Conference on High
Energy Physics}, Chicago and Batavia, Illinois, Sept. 6 -- 13, 1972,
edited by J. D. Jackson, A. Roberts, and R. Donaldson (Fermilab, Batavia,
IL, 1972)}
\def \ijmpa#1#2#3{Int.\ J.\ Mod.\ Phys.\ A {\bf#1}, #2 (#3)}
\def \ite{{\it et al.}}
\def \jhep#1#2#3{JHEP {\bf#1}, #2 (#3)}
\def \jpb#1#2#3{J.\ Phys.\ B {\bf#1}, #2 (#3)}
\def \lg{{\it Proceedings of the XIXth International Symposium on
Lepton and Photon Interactions,} Stanford, California, August 9--14 1999,
edited by J. Jaros and M. Peskin (World Scientific, Singapore, 2000)}
\def \lkl87{{\it Selected Topics in Electroweak Interactions} (Proceedings of
the Second Lake Louise Institute on New Frontiers in Particle Physics, 15 --
21 February, 1987), edited by J. M. Cameron \ite~(World Scientific, 
Singapore, 1987)}
\def \kdvs#1#2#3{{Kong.\ Danske Vid.\ Selsk., Matt-fys.\ Medd.} {\bf #1},
No.\ #2 (#3)}
\def \ky85{{\it Proceedings of the International Symposium on Lepton and
Photon Interactions at High Energy,} Kyoto, Aug.~19-24, 1985, edited by M.
Konuma and K. Takahashi (Kyoto Univ., Kyoto, 1985)}
\def \mpla#1#2#3{Mod.\ Phys.\ Lett.\ A {\bf#1}, #2 (#3)}
\def \nat#1#2#3{Nature {\bf#1}, #2 (#3)}
\def \nc#1#2#3{Nuovo Cim.\ {\bf#1}, #2 (#3)}
\def \nima#1#2#3{Nucl.\ Instr.\ Meth. A {\bf#1}, #2 (#3)}
\def \np#1#2#3{Nucl.\ Phys.\ {\bf#1}, #2 (#3)}
\def \PDG{Particle Data Group, K. Hagiwara {\it et al.},
Phys.~Rev.~D {\bf 66}, 010001 (2002)}
\def \pisma#1#2#3#4{Pis'ma Zh.\ Eksp.\ Teor.\ Fiz.\ {\bf#1}, #2 (#3) [JETP
Lett.\ {\bf#1}, #4 (#3)]}
\def \pl#1#2#3{Phys.\ Lett.\ {\bf#1}, #2 (#3)}
\def \pla#1#2#3{Phys.\ Lett.\ A {\bf#1}, #2 (#3)}
\def \plb#1#2#3{Phys.\ Lett.\ B {\bf#1}, #2 (#3)}
\def \pr#1#2#3{Phys.\ Rev.\ {\bf#1}, #2 (#3)}
\def \prc#1#2#3{Phys.\ Rev.\ C {\bf#1}, #2 (#3)}
\def \prd#1#2#3{Phys.\ Rev.\ D {\bf#1}, #2 (#3)}
\def \prl#1#2#3{Phys.\ Rev.\ Lett.\ {\bf#1}, #2 (#3)}
\def \prp#1#2#3{Phys.\ Rep.\ {\bf#1}, #2 (#3)}
\def \ptp#1#2#3{Prog.\ Theor.\ Phys.\ {\bf#1}, #2 (#3)}
\def \rmp#1#2#3{Rev.\ Mod.\ Phys.\ {\bf#1}, #2 (#3)}
\def \rp#1{~~~~~\ldots\ldots{\rm rp~}{#1}~~~~~}
\def \si90{25th International Conference on High Energy Physics, Singapore,
Aug. 2-8, 1990}
\def \slc87{{\it Proceedings of the Salt Lake City Meeting} (Division of
Particles and Fields, American Physical Society, Salt Lake City, Utah, 1987),
ed. by C. DeTar and J. S. Ball (World Scientific, Singapore, 1987)}
\def \slac89{{\it Proceedings of the XIVth International Symposium on
Lepton and Photon Interactions,} Stanford, California, 1989, edited by M.
Riordan (World Scientific, Singapore, 1990)}
\def \smass82{{\it Proceedings of the 1982 DPF Summer Study on Elementary
Particle Physics and Future Facilities}, Snowmass, Colorado, edited by R.
Donaldson, R. Gustafson, and F. Paige (World Scientific, Singapore, 1982)}
\def \smass90{{\it Research Directions for the Decade} (Proceedings of the
1990 Summer Study on High Energy Physics, June 25--July 13, Snowmass, 
Colorado),
edited by E. L. Berger (World Scientific, Singapore, 1992)}
\def \tasi{{\it Testing the Standard Model} (Proceedings of the 1990
Theoretical Advanced Study Institute in Elementary Particle Physics, Boulder,
Colorado, 3--27 June, 1990), edited by M. Cveti\v{c} and P. Langacker
(World Scientific, Singapore, 1991)}
\def \yaf#1#2#3#4{Yad.\ Fiz.\ {\bf#1}, #2 (#3) [Sov.\ J.\ Nucl.\ Phys.\
{\bf #1}, #4 (#3)]}
\def \zhetf#1#2#3#4#5#6{Zh.\ Eksp.\ Teor.\ Fiz.\ {\bf #1}, #2 (#3) [Sov.\
Phys.\ - JETP {\bf #4}, #5 (#6)]}
\def \zpc#1#2#3{Zeit.\ Phys.\ C {\bf#1}, #2 (#3)}
\def \zpd#1#2#3{Zeit.\ Phys.\ D {\bf#1}, #2 (#3)}

\end{document}